\documentclass[a4paper,11pt,twoside]{article}

\usepackage{graphicx} 
\usepackage{amsmath}
\usepackage{amssymb}

\usepackage[abbr]{harvard}
\citationstyle{dcu}
\citationmode{abbr}
\renewcommand{\harvardand}{\&}

\newcommand{\oldstyle}[1]{#1}
\newcommand{\at}{{\char '100}}
   
\newcommand{\DEF}{\stackrel{\mathrm{def}}{=}}
\newcommand{\DEFt}{\stackrel{\mbox{\rm\tiny def}}{=}}

\newcommand{\ket}[1]{|#1\rangle} 
\newcommand{\kete}[1]{|\kern.3ex#1\kern.3ex\rangle}
\newcommand{\bra}[1]{\langle #1 |}
\newcommand{\brae}[1]{\langle\kern.3ex #1 \kern.3ex|} 
\begin{document}

\title{A differential method for bounding the ground state energy }
\author{Amaury Mouchet\thanks{mouchet\at phys.univ-tours.fr}}
 
\date{Laboratoire de Math\'ematiques 
                            et de Physique Th\'eorique \\
                            \textsc{(cnrs umr 6083)},\\ Universit\'e Fran\c{c}ois Rabelais
                            Ave\-nue Monge, Parc de Grandmont 37200
  Tours, 
                            France.} 

\maketitle

\begin{abstract} 
For a wide class of Hamiltonians, a novel method to obtain lower and
upper bounds for the lowest energy is presented. Unlike perturbative
or variational techniques, this method does not involve the
computation of any integral (a normalisation factor or a matrix
element). It just requires the determination of the absolute minimum
and maximum in the whole configuration space of the local energy
associated with a normalisable trial function (the calculation of the
norm is not needed). After a general introduction, the method is
applied to three non-integrable systems: the asymmetric annular
billiard, the many-body spinless Coulombian problem, the hydrogen atom
in a constant and uniform magnetic field.  Being more sensitive than
the variational methods to any local perturbation of the trial
function, this method can used to 
systematically improve the energy bounds with a
local skilled analysis; an algorithm relying on this method can
therefore be constructed and an explicit example for a one-dimensional
problem is given.
\end{abstract}

{PACS : 03.65.Db, 05.45.Mt, 02.30.Tb, 02.60.Gf.}

\section{Introduction}

In a large variety of interesting physical problems, finding the
discrete spectrum of an operator can be done with approximate methods
only.  Moreover, in most cases, it is a rather difficult task to
estimate errors. Perturbative techniques often lead to non-convergent
series and an evaluation of the discrepancies with the exact result is
usually beyond their scope.  For semibounded operators, variational
methods naturally provide upper bounds for the lowest eigenvalue and
require much more work for providing a lower bound with Temple-like
methods \cite[XIII.2]{Reed/Simon78a}.  Another source of difficulties
when dealing with perturbative and/or variational techniques, is that
they both involve the calculation of integrals on the configuration
space $\mathcal{Q}$: the norm of the wavefunctions and some matrix
elements of operators. This reflects the fact that the discrete
spectrum of a differential operator encapsulates some global
information on $\mathcal{Q}$ within the boundary conditions imposed on
the normalisable wavefunctions. In this article, I want to propose an
approximate method that will overcome these two obstacles: it can
rigorously provide both lower and upper bounds without any kind of
integration.  Like the variational techniques, it will involve a (set
of) trial normalisable function(s) with the appropriate boundary
conditions and will concern in practice the lowest eigenvalue only.
The bounds are given by the absolute extrema of a function defined on
$\mathcal{Q}$ (the so-called local energy). In a sense, this method
allows to stay as local as possible in the configuration space
$\mathcal{Q}$: in order to improve the bounds, a local analysis near
the extrema (or near the possible singular points) of the local energy
is sufficient and necessary.  
This paper is organised as follows. In
section~\ref{sec:bounds}, I give the proof of the inequalities
that will be the  starting point of the differential method.
A comparison with what already exists in the litterature follows
and the guidelines of the method are presented.
Sections~\ref{sec:billiards}, \ref{sec:manybodycoulomb} and \ref{sec:HinB}
show how the method can be applied to three non-integrable quantum systems. 
 Before the concluding remarks, in
section~\ref{sec:algorithm}, I explicitely show on the quartic oscillator
how the  sensitivity of the differential method to local perturbations of the
trial functions can be exploited to systematically improve the bounds
of the ground state energy with an elementary algorithm.

\section{Bounding the ground state energy with the local energy}
\label{sec:bounds}

Let us start with  a quantum system whose
Hamiltonian $\hat{H}$ acts on the Hilbert space of
functions defined on a configuration space $\mathcal{Q}$. Let us
suppose that $\hat{H}$ has an eigenstate 
$\ket{\Phi_0}$  associated with an element $E_0$ of the discrete spectrum.
For any state $\ket{\varphi}$, the hermiticity
of $\hat{H}$ implies the
identity $\bra{\Phi_0}(\hat{H}-E_0)\ket{\varphi}=0$. If we
choose $\ket{\varphi}$ such that its configuration space
representation $\varphi(q)$ is a smooth real normalisable
wavefunction, we obtain:
\begin{equation}
\label{eq:integralidentity}
        \int_\mathcal{Q}\Phi^*_0(q)(H-E_0)\varphi(q)\,dq=0\;.
\end{equation}
The crucial \emph{positivity hypothesis} is to assume that we can choose one eigenstate
such that its eigenfunction $\Phi_0$ remains real and positive or zero in the
whole $\mathcal{Q}$.  This generally applies to the ground state for
which it has been shown in many cases that it is 
strictly positive in the interior of $\mathcal{Q}$
\cite[XIII.12]{Reed/Simon78a}. 
  Then, the real part
of \eqref{eq:integralidentity} involves an integrand that is a smooth,
real function constructed from the real part $H_{\!R}$ of the
differential operator $H$.  Then there exists a $q$ in $\mathcal{Q}$ such
that $\Phi_0(q)(H_{\!R}-E_0)\varphi(q)$ changes its sign. Therefore,
\begin{equation}
        \label{eq:Hectoridentity}
        \exists\, q\in \mathcal{Q}\quad \text{such that}\quad(H_{\!R}-E_0)\varphi(q)=0\;. 
\end{equation}
 Let us now introduce a function on $\mathcal{Q}$ that is known as the
\textit{local energy} \footnote{The usual motivation for introducing the local
energy is just to roughly estimate the dispersion in energy obtained
for an approximated eigenfunction. For instance, when using Monte-Carlo
methods for computing expectation values.
The derivation of inequality \eqref{eq:inequality} shows that for
the ground state this qualitative approach can be made rigourous.}
\begin{equation}
        \label{def:localenergy}
        E_\mathrm{loc}^{[\varphi]}(q)\ \DEF\ \frac{H_{\!R}\,\varphi(q)}{\varphi(q)}\;.
\end{equation} 
From condition \eqref{eq:Hectoridentity}, we immediately obtain that for all smooth real 
and normalisable state $\varphi$,
\begin{equation}
        \label{eq:inequality}
        \inf_\mathcal{Q} \big(E_\mathrm{loc}^{[\varphi]}(q)\big)
        \leqslant\ 
        E_0\ \leqslant\sup_\mathcal{Q} \big(E_\mathrm{loc}^{[\varphi]}(q)\big)\;.
\end{equation}
Surprisingly, these two inequalities are sparsely known in the
literature
\cite{Barnsley78a,Baumgartner79a,Thirring79b,Crandall/Reno82a,Schmutz85a}
and always under some more restricted conditions (the upper bound is
often missing).

The original proof presented here links the inequalities to the
non-ne\-ga\-ti\-vi\-ty of the ground state without referring to the detailed
structure of the Hamiltonian.  In particular, it does not require for
the Hamiltonian to have the purely quadratic form
\begin{equation}\label{eq:SchroHam}
  \hat{H}=\sum_{i,j}a_{i,j}\hat{p}_i\hat{p}_j
+V(\hat{q})
\end{equation}
 where $a$ is a definite positive real matrix and~$V$ a well-behaved
potential.  Inequalities~\eqref{eq:inequality} still apply (with the
appropriate definition \eqref{def:localenergy} of the local energy) in
the presence of a singular potential, when there is a magnetic field
and for an infinite number of freedoms (like in the non-relativistic
quantum field describing a BEC condensate).  Besides, it is not
required for $\varphi$ to be nonvanishing.  It simply says that where
$\varphi$ vanishes faster than $H_{\!R}\,\varphi$, one or both of the
bounds can be infinite and therefore useless.

  The first form of inequalities \eqref{eq:inequality} (with its two
bounds) is due to Barta \cite{Barta37a} and was derived for the
fundamental vibration mode of an elastic membrane. Though Barta writes
that his method will be generalised in subsequent publications, I was
not able to find any extensions of his original work before an article
of Duffin \cite{Duffin47a} where a Schr\"odinger operator of the form
$H=-\Delta +V$ is considered. Duffin shows that the Dirichlet boundary
conditions imposed on the trial function $\varphi$ can be relaxed but
he loses the upper bound. One obtains the equalities in
\eqref{eq:inequality} for a flat local energy i.e. for
$\varphi=\Phi_0$; hence we will try to work with a $\varphi$ that
mimics the exact ground state best. Therefore, generalising the Barta
inequalities by increasing the size of the functional space of
$\varphi$'s can be irrelevant. One should instead keep working with a
restricted set of trial functions that respects some a priori known
properties of $\Phi_0$, such as its boundary conditions, its
symmetries and its 
positivity\footnote{A technicality should be mentioned here: If one
  chooses the trial states~$\ket{\varphi}$ such that for all $q\in\mathcal{Q}$,
  $\langle q \ket{\varphi}>0$ except, perhaps, for the a priori known
  zeros of~$\langle q \ket{\Phi_0}$, we can deal with systems where the configuration
  variable~$q$ includes some discrete parameter like a spin index (the
  somewhat loosely notation $H_{\!R}\varphi(q)$ must be understood as
  the real part of 
  $\bra{q}\hat{H}\ket{\varphi}$ and $dq$ is the measure on $\mathcal{Q}$ possibly having a 
  continuous and/or a discrete part). Indeed, under the positivity
  hypothesis, from
  \eqref{eq:integralidentity} we deduce that there must be a
  couple~$(q,q')$ in $\mathcal{Q}^2$ such that $\varphi(q)>0$,
  $\varphi(q')>0$, $(H_{\!R}-E_0)\varphi(q)\geqslant0$,
  $(H_{\!R}-E_0)\varphi(q')\leqslant0$.  Therefore
  inequalities \eqref{eq:inequality} remain valid. For instance, if
  $H_{\!R}$ is a (possibly finite) matrix, the local energy consists of a
  discrete (finite) set of real numbers.}.

More precisely, we will explain in the last part of this
paper that, once a $\varphi$ that bounds the local energy is found, it
is expected that there is only a \textit{finite} number of independent
directions in the functional space along which the bounds can be
improved.  In the following we will actually deal with a finite
dimensional submanifold of trial functions $\varphi_\lambda$ where
$\lambda$ stands for a small number of control parameters varying in a
control space $\mathcal{C}$. Accordingly, the strategy is clear: for,
say, obtaining an optimized lower bound we will try to find
$\sup_{\lambda\in\mathcal{C}}\left( \inf_{q\in\mathcal{Q}}
\big(E_\mathrm{loc}(\lambda;q)\big)\right)$ where
$E_\mathrm{loc}(\lambda;q)$ stands for
$E_\mathrm{loc}^{[\varphi_\lambda]}(q)$.  As long as the extremal
values of $q\mapsto E_\mathrm{loc}(\lambda;q)$ can be followed
smoothly with $\lambda$ (in particular the Morse points are
generically stable), the problem is reduced to local differential
calculations in $\mathcal{C}\times\mathcal{Q}$ in the neighborhood of
the critical points: adding to a trial function an infinitesimal
perturbation that is localized far away from the extremal point does
not affect the energy bounds.  One recovers the global sensitivity of
the eigenvalue problem because the critical points of the local energy
generically bifurcate for finite variations of $\lambda$
\cite{Poston/Stewart78a,Demazure00a} and can jump to other distant
points when a degeneracy occurs \cite[chap. 10, especially
fig. 50]{Arnold84a}.

In mathematical physics literature, Barta's inequalities are always
considered within the context of the billiards systems (Laplacian
spectra on a Riemannian manifold), even in the most recent papers (for
instance \cite{Pacelli/Montenegro04a}).  As far as I could search, the
most advanced extension to other physical problems has been made
(tentatively) by Barnsley \cite{Barnsley78a} but, for the same reasons
as Duffin's \cite{Duffin47a}, he systematically loses the upper bound.
Besides, he acknowledges he  is unable to produce any non-trivial bound for the Helium
atom.

One can easily understand Barnsley's failure: With variational
methods, a very rough estimation of the exact ground state
wavefunction can lead to a reasonably good agreement for $E_0$ while a
simple local perturbation of the exact wave function can even make the
local energy unbounded.  Therefore, at first sight, one could see the
sensitivity of the local energy as a major drawback of the method:
variational methods are more robust to local perturbations of the
trial function.
  But this argument can be reversed : compared to the
rigidity of the variational methods, the differential method offers
the possibility to improve  the estimations at low cost provided
we are able to implement a skilled strategy (eliminating the
singularities, controlling the behavior at infinity with \textsc{jwkb}
techniques, increasing the absolute minimums, etc.).  In the
following, I explicitly show in many non-trivial cases, that once we
have this strategy in mind, we can obtain interesting results
for complex systems.  For instance, not only we can improve Barnsley's
trivial bound for the ground state energy of the Helium atom, but it
will be shown in section~\ref{sec:manybodycoulomb} how this result
generalizes to any number of Coulombian particles.  The calculations
can be made analytically with a surprising simplicity.

As far as the upper bound is concerned, the variational method
leads \textit{a priori}  to a better approximation than the
differential method since, for any normalised function~$\phi$,
\begin{equation}
  \int_\mathcal{Q}\varphi^*(q)H\varphi(q)\,dq
=\int_\mathcal{Q}|\varphi(q)|^2 E_\mathrm{loc}^{[\varphi]}(q)\,dq
\leqslant \sup_\mathcal{Q} \big(E_\mathrm{loc}^{[\varphi]}(q)\big)\;.
\end{equation}
Nevertheless, being free of any integration, 
the absolute maximum of the local energy is a quantity that 
is more easily accessible to analytical or numerical computations than the
average value of~$H$.
 
\section{Application to billiards; the example of the 2d-annular billiard}
\label{sec:billiards}

As a first illustration of the differential method, let us consider
the problem of finding the lowest eigenvalue of $H=-\Delta/2$ in a
connected finite region $\mathcal{Q}$ with the Dirichlet boundary conditions
imposed on $\mathcal{B}\DEFt\partial\mathcal{Q}$. Suppose that the
boundary $\mathcal{B}$ is given by an implicit smooth scalar equation
of the form $b(q)=0$ while the interior of $\mathcal{Q}$ is defined to
be the set of points $q$ such that $b(q)<0$. Then, trial functions can
be taken of the form $\varphi=fb$ for any arbitrary smooth function $f$
that does not vanish inside $\mathcal{Q}$. The only possible singular
points of $q\mapsto E_\mathrm{loc}^{[\varphi]}(q)$ are located
on $\mathcal{B}$ and can be removed if $f$ is chosen with an
appropriate behaviour in the neighborhood of $\mathcal{B}$.  Imposing
this behaviour for $f$ is a priori a simpler task than solving the
eigenvalue problem on the global $\mathcal{Q}$: one dimension has been
spared since we have to deal with some local properties of $f$
near $\mathcal{B}$. For instance, by generalizing Barta's trick
\cite{Barta37a}, one can easily check by a simple equation counting
that when $\mathcal{Q}$ is algebraic i.e. when $b$ is a polynomial,
provided that we choose $f$ to be a polynomial of sufficiently high
degree $n$ whose zeros are outside $\mathcal{Q}$, we can find a
polynomial $g$ of degree $n-2$ such
that $\Delta(fb)=gb$. Therefore $E_\mathrm{loc}^{[\varphi]}=-g/(2f)$ is
bounded and finite upper and lower bounds of $E_0$ can be
found.  Let us apply this method to the asymmetric annular billiard
that is an elegant paradigmatic model in quantum chaos
\cite{Bohigas+93a}. $\mathcal{B}$ is made of two circles of radius $1$
and $r<1$ whose centers are distant by $\delta<1-r$. $\mathcal{Q}$ is
the 2d-domain in between the circles.  The simplest choice of trial
function is to take
$\varphi(x,y)=b(x,y)=[x^2+y^2-r^2][(x-\delta)^2+y^2-1]$. One
 can check analytically that the lower bound of
$E_\mathrm{loc}^{[\varphi]}$ is 
\begin{equation}\label{eq:Elocbilliard}
        \inf_\mathcal{Q}\left(\frac{-\Delta\varphi}{2\varphi}\right)
        =\sup_\mathcal{Q}\left(
          \frac{8[(x-\delta/2)^2+y^2-(1+r^2)/4]}
                {[x^2+y^2-r^2][(x-\delta)^2+y^2-1]}
                         \right).       
\end{equation}
For $r=3/4$
and $\delta=0.1$, a simple numerical computation shows that \eqref{eq:Elocbilliard} is finite and obtained
at $(x,y)\simeq(0.86,0)$ and leads to $E_0\geqslant28.390$ to be
compared with the exact result $E_0\simeq42.94$.  As one could have
expected with the rough trial function chosen above, the estimation is
not very precise but the calculations required here to get this result
are  much simpler than the ones involved in a variational method
(that provides the complementary upper bound $55.32$ with the
same test function) 
or by the exact numerical resolution that requires to find the
smallest root of an infinite determinant made of Bessel functions.

 \section{The many-body Coulombian problem}
\label{sec:manybodycoulomb}

The next examples, presented in this section and in the following, 
will illustrate that the first strategy for
obtaining finite bounds is to get rid of the singularities that may
appear in the local energy. When $\mathcal{Q}$ is not  bounded,
 one must have a control over the behaviour of the trial
functions as $q$ goes to infinity. For a
multidimensional, non separable, Schr\"odinger Hamiltonian, a
\textsc{jwkb}-like asymptotic expression is generally not available
\cite[Introduction]{Maslov/Fedoriuk81a}. Nevertheless, the differential method is
less demanding than the semiclassical approximations: we will ask that
the
local energy be bounded at infinity but we will not require it
to tend to the \textit{same} limit in all directions.  As already shown in the
annular billiard problem, for the sake of simplicity one could start
with a less ambitious program and try to obtain just one nontrivial
inequality in \eqref{eq:inequality}.

The second example is to consider a system of $N$ non-relativistic,
spinless, charged particles living in 
a  $D$-dimensional infinite space. 
Their kinetic energy is given by $\sum_{i=0\dots
N-1}\hat{p}_i^2/(2m_i)$ and they interact with each other via a
two-body Coulombian interaction $e_ie_j/\hat{r}_{ij}$. We will assume that
the masses $m_i$ and the charges $e_i$ allow the existence of a bound
state. Once the free motion of the center of mass is discarded, we are
led to a $D(N-1)$-dimensional configuration space that can be described by the
relative positions $q=\{\mathbf{r}_{0,i}\}_{i=1\dots N-1}$ with
respect to one distinguished particle.  The Hamiltonian is given by
\begin{equation}
        \label{eq:hamNbody}
        \hat{H}=\sum_{i=1}^{N-1}\frac{1}{2m_{0,i}}\hat{\mathbf{p}}_i^2
                +\frac{1}{2m_0}\sum_{\substack{i,j=1\\i\neq j}}^{N-1}\hat{\mathbf{p}}_i.\hat{\mathbf{p}}_j
                +\frac{1}{2}\sum_{\substack{i,j=0\\i\neq j}}^{N-1}\frac{e_ie_j}{\hat{r}_{i,j}}\;.
\end{equation}  
The notation $m_{i,j}$ stands for the reduced mass $m_im_j/(m_i+m_j)$.
For $D\geqslant2$, one can eliminate the Coulombian simple poles
$r_{i,j}=||\mathbf{r}_{0,i}-\mathbf{r}_{0,j}||=0$ in the local energy
by choosing the trial function as follows:
\begin{equation}\label{eq:trialfunctionNbody}
        \varphi(q)=\exp\Big(-\frac{1}{2}\sum_{\substack{i,j=0\\i\neq j}}^{N-1}\lambda_{i,j}r_{i,j}
                       \Big)
\end{equation}
with $\lambda_{i,j}=-2m_{i,j}e_i e_j/(D-1)$. When this choice does not
provide a normalisable function, it should be understood that the
exponent is just the first order of a Taylor expansion near
$r_{i,j}=0$.  Whenever \eqref{eq:trialfunctionNbody} is actually
square integrable on $\mathcal{Q}$, the local energy reads
\begin{equation}
        \label{eq:ElocNbody}
        E_\mathrm{loc}^{[\varphi]}=-\sum_{\substack{i,j=0\\i<j}}^{N-1}\frac{\lambda_{i,j}^2}{2m_{i,j}}
        -
        \sum_{\widehat{j,i,k}}\frac{\lambda_{i,j}\lambda_{i,k}}{m_i}\cos(\widehat{j,i,k}).
\end{equation}
The last sum involves all the angles  $\widehat{j,i,k}$ that can be formed
with all the triangles made of three distincts particles.
  This expression treats all the particles on an equal footing
and it is clear that the local energy is bounded
everywhere. 
Bounding simultaneously all the $\cos(\widehat{j,i,k})$
by $\pm1$ can be a rather crude approximation. One should instead take into
account the correlation between the angles. For instance, when $D=3$, as soon
as $N\geqslant3$, their number ($N(N-1)(N-2)/2$) exceeds the number of
independent variables minus three Euler angles and a dilatation factor
 $3(N-1)-3-1$).
 For $N=2$, we recover the exact ground state since
the last sum is absent and the local energy is
constant. For a helium like atom ($D=3$, $N=3$) of charge $(Z-2)e$
with the nucleus considered as infinitely massive compared to the
electron mass, only two angular terms survive and we get (in atomic
units) $E_\mathrm{loc}=-Z^2-1/4+Z(\cos\theta_1+\cos\theta_2)/2$.  The
two angles are taken at the vertices made by the two electrons:
the sum of their cosines is bound from below by 0 and from above by 2
(diametrically opposed electrons).  The lower bound of the local energy
is $-Z^2-1/4$; this is not an interesting piece of information since we 
know that $E_0$ is larger than the energy $-Z^2$ obtained by neglecting the
strictly positive repulsion of the electrons. On the other hand the
bound $E_0\leqslant-(Z-1/2)^2$ provides a simple, analytical, non-trivial
result.

\section{The hydrogen atom in a magnetic field}
\label{sec:HinB}

\begin{figure}[!ht]
\center  
\includegraphics[width=12cm]{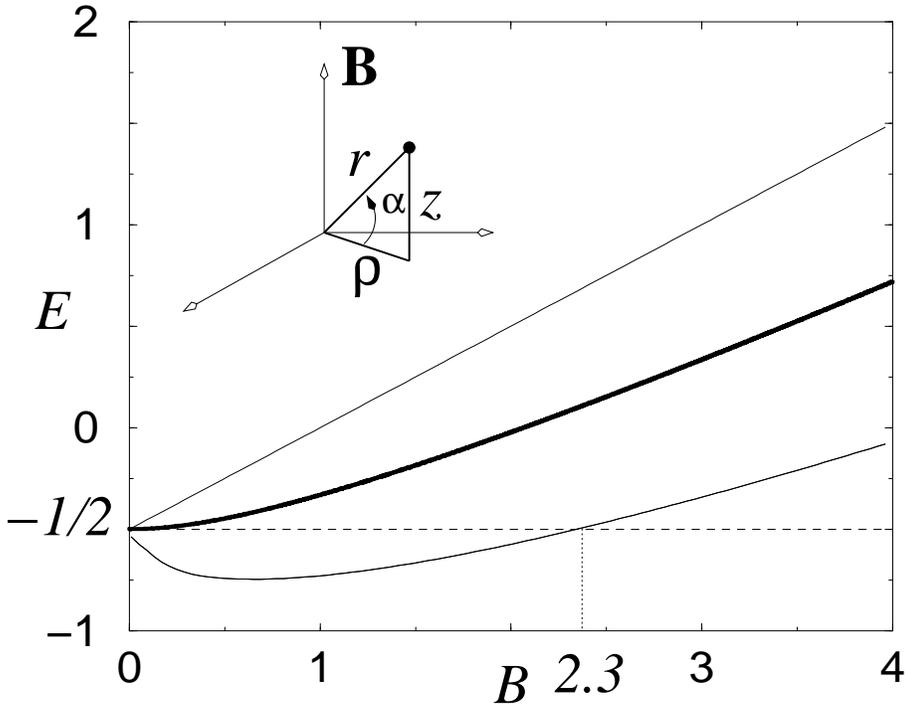}
\caption{\label{fig:HB}Upper and lower bounds of the ground state energy (thick line) for the hydrogen in a Zeeman configuration.}
\end{figure}

The third application of the differential method to a non-integrable
system will concern the hydrogen atom in a constant and uniform
magnetic field $B\vec{u}_z$.  While the positivity of the
ground state wavefunction was guaranteed by the so-called Krein-Rutman theorem
 in all the previous
examples (applicable for any Hamiltonian of the
form~\eqref{eq:SchroHam}, see for instance \cite[XIII.12]{Reed/Simon78a}),
 it is not valid any longer for an arbitrary potential when a
magnetic field is present \cite{Helffer+99a}.  Nevertheless, for the hydrogen
atom, the attractive interaction between the nucleus and the electron
keeps  the orbital momentum $L_z$ of the ground state at zero for any
arbitrary value of $B$ \cite{Avron+77a} and the Krein-Rutman theorem
applies when restricted to the $L_z=0$
subspace. In
order to preserve the symmetry of the ground state, the trial function 
will be chosen to be 
strictly positive and even with respect to $z$;
hence we can work in the half space where $z\geqslant0$.  
 With a vanishing paramagnetic term, $\varphi$
depends on the coordinates $q=(\rho,r)$ only (see FIG.~\ref{fig:HB}).
 In atomic units, the local energy is given by
$E_\mathrm{loc}^{[\varphi]}=V-\Delta\,\varphi/(2\varphi)$ where the
effective potential is $V(\rho,r)=B^2\rho^2/8-1/r$.
 In order to eliminate the Coulomb singularity, one
must impose some local conditions on the logarithmic derivatives
of $\varphi$. More precisely, with $S\DEFt\ln\varphi$, we must have
$\partial_r S(0,0)=-1$ and $\partial_\rho S(0,r)=0$ for
all $r\geqslant0$. Assuming that $S$ is smooth enough near $\rho=0$, it takes 
the general form $S(\rho,r)=-r+r^2l(r)+\rho^2h(\rho,r)$ where
$l$ and $h$ are two smooth functions. Choosing $l\equiv0$
and $h\equiv-B/4$ [resp. $h\equiv0$] will bound from above
[resp. below] the local energy:
$-1/2\leqslant E_0\leqslant-1/2+B/2$. Like in the previous example
the lower bound is useless since it can be guessed from the very
beginning. The real challenge here is to improve the lower bound
without introducing a divergence as $r\to \infty$ in any direction
characterized by $\alpha$. After a detailed examination of
the possible balance between the asymptotic behaviour of $l$ and $h$
as $r\to \infty$, a trial function can be constructed in order to
improve the trivial lower bound for $B$ large enough. Namely if we
take $S=-r-B\rho^2/4+\rho^2(r-\sqrt{r^2-\rho^2})/(\rho^2+5r/\sqrt{B})$ we
improve the trivial lower bound for $B\gtrsim2.3$ (see FIG.\ref{fig:HB}).
Note that for very large $B$ we recover a wavefunction that mimics
a Landau state.

\section{A local algorithm for improving the bounds ; application to the quartic oscillator }
\label{sec:algorithm}

\begin{figure}[!ht]
\center  
\includegraphics[width=12cm]{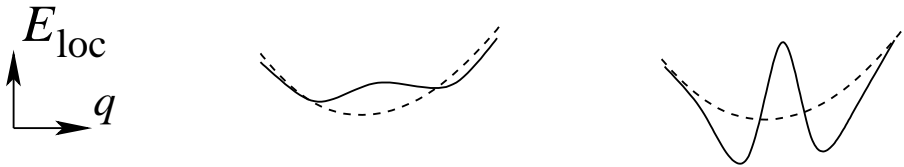}
\caption{\label{fig:censeur}Increasing $E_\mathrm{loc}^{[\varphi_0]}$ (dashed line)
 with one local perturbation is  limited. When adding to $\ln(\varphi_0)$, say, 
a gaussian perturbation localised 
near an isolated  minimum of the local energy, we can increase the value of the absolute minimum
by a finite amount (solid line in the left figure). But if the magnitude of the
Gaussian is increased too much, the minimums created by passing through 
a bifurcation can decrease below the original minimum (solid line in
the right figure).
}
\end{figure}

Assume now that $\varphi_0=\exp(S_0)$ bounds the local energy. Is
there any systematic strategy to improve the bounds and, one day,
compete with the very high precision of the secular variational and
perturbative methods or numerical diagonalisation of truncated
matrices ?  Of course one can always combine these four approaches but
let us look, for the moment, how the local character of the
differential method can be exploited further.  Suppose that $q_0$ is a
point where $q\mapsto E_\mathrm{loc}^{[\varphi_0]}(q)$ reaches its
lowest non-degenerate value.  Among all the possible infinitesimal
perturbations of $S_0$, only those that are localised in the
neighbourhood of $q_0$ are relevant since adding a perturbation far
away from $q_0$ will not affect the absolute minimum. The appropriate
framework for local studies in an infinite functional space is
bifurcation theory. Since the local energy, the determination of the
critical points and their stability involve a finite number of
derivatives, we expect that the number of relevant control
parameters~$\lambda$ remains finite for low-dimension configuration
space very much like the central result of catastrophe theory
\cite{Poston/Stewart78a,Demazure00a}.  We will leave this quantitative study for future
investigations.  For the moment, let us keep the discussion at a
qualitative level only with a 1d Hamiltonian of the form
$H=-\Delta/2+V$ and take $S=S_0+\delta S$ with a Gaussian perturbation
$\delta S(q)=s\exp(-(q-a)^2/\sigma^2)$ controlled by three parameters
$\lambda=(s,a,\sigma)$. The specific choice of the form of~$\delta S$ is not important
here ; only a finite number of pointwise derivatives will matter for locally improving the 
bounds as long as  
  $\delta S$ does not change the
normalisability of the trial function. The choice of a gaussian is particularly simple: it 
 will modify the local energy
in a neighbourhood of~$a$ whose size is governed by~$\sigma$ and by
the magnitude~$s$.  
This perturbation is qualitatively reproduced in
FIG.\ref{fig:censeur}.  The value of $E_\mathrm{loc}$ at~$q=a$ is
increased (resp. decreased) for a small but finite positive
(resp. negative)~$s$. We can apply this procedure near the absolute
minimum (resp. maximum) of~$E_\mathrm{loc}$ and repeat it for the
possible absolute extrema that may have emerged during the previous
step. We get an iteration sequence that may systematically improve the
bounds. Still, this algorithm is slowed down because if we try to
``lift up the dress'' too much, a ``prudish censor'' lowers it on the
both sides of~$q=a$ (this phenomenon is not specific to 1d).

 To be more precise, consider a quartic potential given
 by~$V(q)=r^2q^2(q^2+\eta\, \delta^2)/2$
where $\eta=\pm1$. In order to bound the local energy as~$|q|\to\infty$, 
we can use a \textsc{jwkb}-like expansion for~$S_0$. If we want a uniformly smooth
 expression, we can
take 
\begin{equation}
  \begin{split}
  S_0(q)=-\frac{1}{3}r(q^2+\delta^2)^{3/2}+\frac{1}{2}r\delta^2(1-\eta)(q^2+\delta^2)^{1/2}\\
-\frac{1}{2}\ln(q^2+\delta^2)-\frac{1}{2}r\delta^4(q^2+\delta^2)^{-1/2}\;.
\end{split}
\end{equation}
\begin{figure}[!ht]
\center  
\includegraphics[width=12cm]{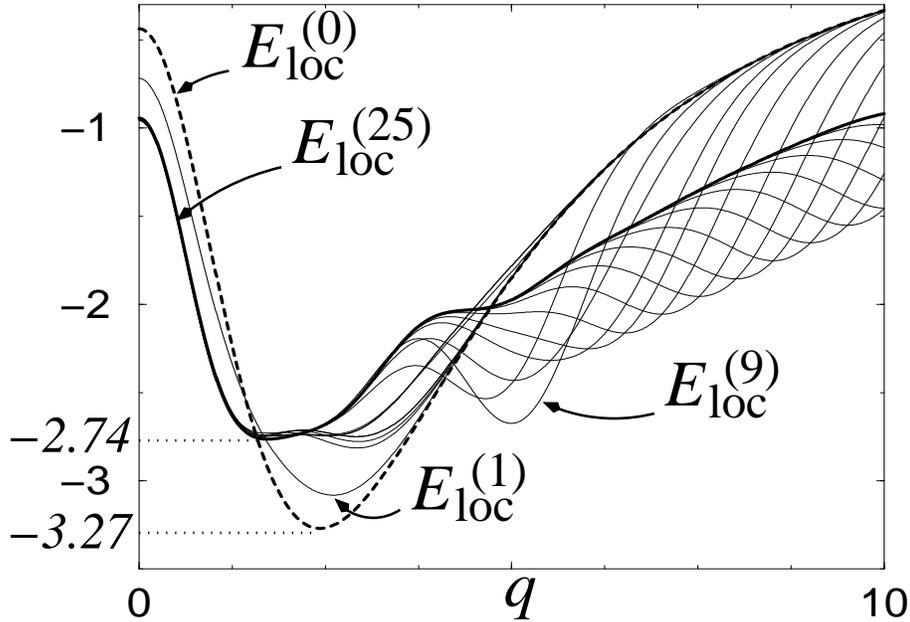}
\caption{\label{fig:Elocq4} Adding~$n$ Gaussians to~$S_0$ allows to increase the minimum 
of the local energy when constructing the sequence~$E_\mathrm{loc}^{(n)}$. For each iteration, only
 one scalar parameter is optimized (the amplitude of the gaussian being added).  }
\end{figure}
The last term is chosen to improve the trivial lower bound given by
the minimum of~$V$.
For the arbitrary choice $r=1/\sqrt{2}$, $\eta=-1$ and $\delta^2=8$, FIG.~\ref{fig:Elocq4} shows how the lower bound can be 
improved from~$-3.27$ up to~$-2.74$ (the exact result is -2.66) when adding to~$S_0$ 
enough Gaussians that are equi-spaced by~$.5$ with a fixed~$\sigma=1$.
Only their magnitude~$s$ are numerically optimized here, one after
another.  One can see a second advantage of the
differential
method when numerically implemented: not only no integral is required
but also, provided we keep under control the instabilities that are illustrated 
in FIG.~\ref{fig:censeur}, 
the optimization algorithm concerns a small number of parameters
at each step (just one in the example given in FIG.~\ref{fig:Elocq4})
to be compared with the large number of parameters to be 
optimized at one go in the final step of the variationnal method.

\section{Conclusion}
\label{sec:conclusion}

The differential method appears to be new kind of general theoretical
tool for obtaining rigourous information on a ground state energy. Its
local character makes it quite different from the traditional ones (to
put it succintly, the variational, the perturbation and the numerical
diagonalization techniques). In this paper, I have given some
qualitative and quantitative arguments to show how simple and
efficient it can be. However, I should insist that even in the cases
where the variational or perturbative techniques can be applied, the
aim of the paper is not to seek for performance: for the moment the
differential method is too young to compete by itself with the
traditionnal methods. One short term possibility is to calculate the
extrema of the local energy constructed with the trial function given
by the other methods. The idea of locally modifying the local energy
or any local function of the same type --- for instance, those
currently used in Monte-Carlo methods --- may be fruitful as well
\cite{Caffarel04a}.  Actually, in order to convince the reader that the
method is indeed applicable in a wide field of physics and furnishes
some reasonable results, I had to compare them with some more precise
ones and therefore I dealt with situations where the exact ground
state energy was already known with the help of other methods. The
algorithm that is presented in section~\ref{sec:algorithm} is chosen to prove
how the local sensitivity of the local energy can be exploited to
systematically improve the bounds.  The feasibility is in itself not
obvious and is worth to be demonstrated even in the simplest cases.

This work could not have been started without Hector Giacomini's
brilliant intuition that some relevant information on~$E_0$ could be
extracted from \eqref{eq:Hectoridentity}.  I am very indebted to
Dominique Delande and Beno\^{\i}t Gr\'emaud for sharing their
penetrating thoughts, their skilled numerical calculations in
Coulombian problems and, not least, their kindful hospitality
 at the Laboratoire 
Kastler Brossel.



\ifx\undefined\BySame
\newcommand{\BySame}{\leavevmode\rule[.5ex]{3em}{.5pt}\ }
\fi
\ifx\undefined\textsc
\newcommand{\textsc}[1]{{\sc #1}}
\fi
\ifx\undefined\emph
\newcommand{\emph}[1]{{\em #1\/}}
\fi

\end{document}